\documentclass{article}%
\usepackage{amssymb}
\usepackage{amsfonts}
\usepackage{amsmath}
\usepackage{graphicx}%
\begin{document}

\title{{\LARGE Charges suspended by strings: A new twist to an old problem}}
\author{Peretz D. Partensky \\Graduate Group in Biophysics, University of California, \\San Francisco, San Francisco, CA,  94143.\\Email: ppartens@itsa.ucsf.edu
\and Michael B. Partensky\\Department of Chemistry, and the Rabb School\\of Continuing \ Studies, Brandeis University,\\ Waltham, MA, 02454.\\Email: partensky@gmail.com}
\maketitle

\section{Introduction}

A popular problem \cite{Gia97} of calculating the equilibrium state in a
system of two charged particles suspended on strings conventionally deals with
mutually repelling charges (\textit{Fig.1, a}). The students are typically
asked to find the equilibrium separation $s$ between the charges or a related
quantity such as the deflection angle. In this paper, we are slightly
modifying this problem by considering two opposite and therefore attracting
charges (\textit{Fig.1, b}). Now, in order to generate an equilibrium state at
finite distance $s$, we suspend the charges at a finite initial separation,
rather than from the same point. One might expect that the problem with
opposite charges is just a routine extension of the original one. It turns out
however, that this is not case. The discussion leading to the solutions
introduces a catastrophic behavior typically avoided in high school physics.

\begin{center}%
\begin{figure}

\begin{center}
\includegraphics[
height=3.7282in,
width=2.2943in
]%
{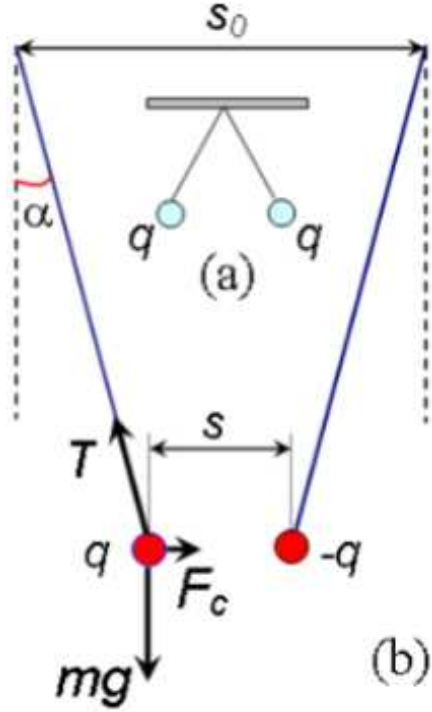}%
\end{center}
\caption{Equilibrium state of two charges
suspended on strings. (a) illustrates the conventional problem with two
identical charges (b) illustrates the modified problem with two opposite
charges suspended at finite separation $s_{0}$ The corresponding force diagram
is shown for $q > 0 $.}
\end{figure}

\end{center}

\section{Problem}

Two identical and oppositely charged$\ $\footnote{Opposite charges on the
conductive balls can be maintained by connecting them to the poles of a
battery, or to the opposite plates of a capacitor.}$\ $particles, $+q$ and
$-q$, are suspended on two strings at an initial separation $s_{0.}$ The mass
of each particle is $m$ and the length of the string is $l$. For simplicity,
assume that $l\ >>s_{0}$ . This means that $\alpha$ can be treated as a small angle.

(A) Find how incrementing charge value $q$ starting from $q=0~$affecs the
distance $s$ between the particles while neglecting the size of the particles.
This is known as a point charge approximation.

(B) Now, view each particle as a point charge surrounded by an insulated hard
spherical shell of radius $R$. How is the equilibrium relation $s(q)$ affected
by R?

(C) Compare the "forward" process (when $q$ is incremented from $q=0)$ with
the "reverse" (discharge) process when $q$ is decremented from the large
values to $q=0$.

\section{Solution}

The equilibrium state of a system is defined by a force balance condition.
There are three forces acting upon each charge: the linear tension of the
string, $T$, the force of gravity \textit{mg} (where $g$ is free acceleration)
and the electrostatic force of attraction $F_{c}$. These forces are
illustrated for one of the charges in \textit{Fig.1b}. The electric force is
determined by Coulomb's law
\begin{equation}
F_{c}=k\frac{q^{2}}{s^{2}} \label{C_Force}%
\end{equation}
where $k$ is the electrostatic constant ($9\cdot10^{9}\ N~m^{2}/C^{2})~$and $s$ is
the separation between the charges. The condition that the net force vanishes
results in the vector equation $\vec{T}+m\vec{g}+\vec{F}_{c}=0$. Splitting
this into horizontal and vertical components yields the equations
$T$sin($\alpha)=F_{c}$ and $T$ cos($\alpha)$ = \textit{mg } Dividing the first
by the second, we get:
\begin{equation}
\tan(\alpha)=\frac{F_{c}}{mg} \label{tang_equilibr}%
\end{equation}

Formally this is identical to the equilibrium condition of the prototype
problem\ \cite{Gia97}. However, if we now rewrite the equation\textit{ }in
terms of the variable $s~$\ features which are particular to our problem are
revealed. To express $\tan(\alpha)$ through $s$, we notice that $\sin
(\alpha)=(s-s_{0})/2l$. \ Recalling that $\alpha$ is a small angle we may
write $\tan(\alpha)\approx\sin(\alpha)=(s-s_{0})/2l$. Using this and
Eq.\ref{C_Force}, \ \textit{Eq.2} can be rewritten as%

\begin{equation}
(s-s_{0})s^{2}=2k\frac{lq^{2}}{mg} \label{s_q}%
\end{equation}

The problem can be solved generally in dimensionless units \footnote{To
simplify the form of \textit{Eq.3}, we can divide both sides of the equation
by $s_{0}^{3}$ and introduce a new \textquotedblleft scaled\textquotedblright%
\ dimensionless distance $z=s/s_{0}~$and charge $\widetilde{q}$ $=q/\gamma
~$with all constants conveniently grouped as $\gamma=\sqrt{mgs_{0}^{3}/2kl~}%
$with units of charge. This handy substitution gives: $z^{2}(1-z)=\widetilde
{q}$ $^{2}$. This equation can be generally solved for any system in
dimensionless units of charge and distance. Details pertaining a specific
system could then obtained by using the numeric parameter values to find
$s_{0}$ and $\gamma~$and rescaling $z(\widetilde{q}$ $)$ back to $s(q)$.} .
However, since it is more traditional to use numerical examples for high
school physics, further analysis we choose $l=1~$m$,~m=10^{-3}~$%
kg$,~s_{0}=0.2~$m. The other constants are $g=9.8$ m/s$^{2}$ and $k=9\cdot10^{9}$ N
m$^{2}$ /C$^{2}$.

\textit{Eq.\ref{s_q}} completely describes the dependence $s(q)~$as requested
in part A of the problem. However, the solution of this equation is not as
trivial as for the prototype case. To illustrate how, we solve
\textit{Eq.\ref{s_q}} graphically by choosing different separations 0 $<
s < s_{0}$ and determining the corresponding $q$ from the equation. A
related problem along with a pedagogical justification of the graphical
approach can be found in \textit{Can a Spring Beat the Charges?}
\cite{ParPar2004}

\begin{center}
\begin{figure}
\begin{center}
\includegraphics[
natheight=0.997100in,
natwidth=1.235000in,
height=2.54in,
width=3.14in
]%
{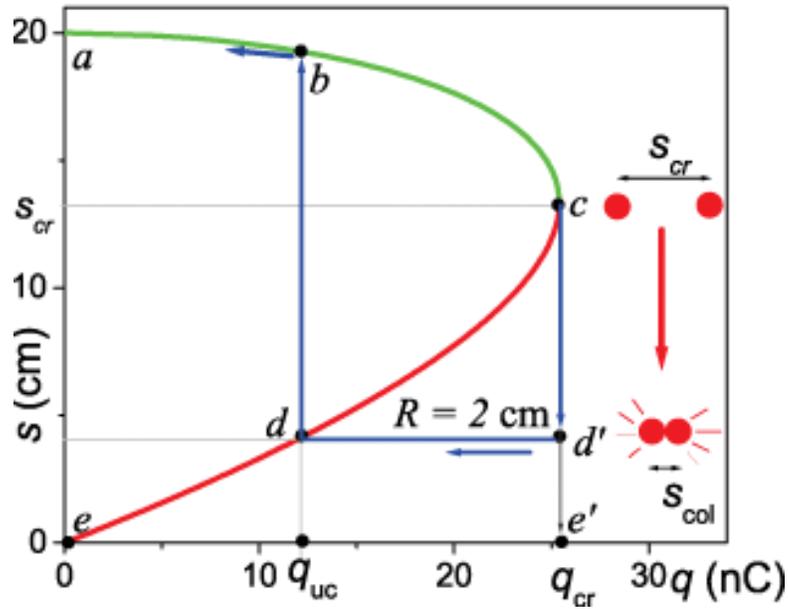}%
\end{center}
\caption{The green and red curves combine to form the plot
of equilibrium distance dependence on charge, $s(q)$. The point where the two
branches meet is labeled $c$ = ($q_{cr},s_{cr})$ and is termed the
\textit{critical point}. Note, $s$ is positioned along the vertical axis which
is the conventional way of presenting $s(q)$ dependence. Points a-e and the
interconnecting paths will be referenced in the discussion.}
\end{figure}

\end{center}

The solution is shown in \textit{Fig.2}. We can now ask, what is the
equilibrium separation for any given value of $q$ ? Surprisingly, for
\textit{0 $<$ q $<$ q}$_{cr}$ the graph gives two possible answers,
represented by the green and red branches of the curve, while for $q$%
$>$%
$q_{cr}$ there are no solutions at all, i.e. $s(q)$ does not penetrate in that
region. $q_{cr}$ denotes the boundary point between two regions where there is
only one solution and s$_{cr}$ is the corresponding separation. Which of two
solutions found in the region $0<q<q_{cr}$ is \textquotedblleft
real\textquotedblright? In the next section it will be explained that these
are the points on the green branch. The smaller equilibrium separations $s$%
$<$%
$s_{cr}$, i.e. the red branch, correspond to points of maxima of the
underlying energy function and are therefore unattainable. We will also
discover that this red branch becomes useful in the analysis of hysteresis (in
answering the question C of the problem).

Let's now examine the physical effects of varying the charge $q$ in closer
detail. As charge $q$ grows from $0$ (the "forward" process), the slope $s(q)$
becomes progressively steeper as demonstrated by the green curve. Finally,
when $q$ approaches the critical value $q_{cr}\mathit{\sim25.4}$\textit{ nC}
the slope of $s(q)$ becomes vertical (see $c$ in \textit{Fig.2}).

What is the significance of this point for our system? At this critical point,
the equilibrium disappears and the charges suddenly \textit{collapse}. The
distance between the charges changes instantly from $13.4$ \textit{cm} (point
$c)$ to 0 (point $e^{\prime})$.

The point charge approximation leads to a non-physical consequence. After the
charges collapse, they stay intact attracted by an infinite force. The only
way to separate them is to reduce $q$ to zero, and thereby entirely cancel the
electrostatic attraction. While the point charge model is a useful
approximation to get a grip on general qualitative behavior, in application to
the collapsed state it becomes unrealistic.

Even the microscopic charges of interest, such as the ions or charged
amino-acids in proteins, have finite size which prevents them from a complete
collapse. To treat the collapsed state and the related behavior more
realistically, we should assign finite sizes to our charges. We avoid some
additional complications, such as discharge through contact\textbf{,} by
assuming each charge is surrounded by a rigid insulating shell of finite
radius R. This was the reasoning behind part B of the problem.

If the spheres have finite radii $R$, their closest possible separation is
$2R$. Consequently, the length of the collapse in such a system decreases from
$s_{cr}$ to $s_{cr}-2R$. As an example, the collapsed state for $R=2$
\textit{cm} is represented in \ \textit{Fig.2} by the blue line $d`d$separated
from point $c$ by a vertical step = $13.4-4$ $=9.4~$\textit{cm}. From this,
one can see that for $R\geq s_{cr}/2$ the catastrophe disappears entirely
because the shells prevent the system from entering the critical range. In
such a case, increasing $q$ leads to continuous displacement of the charges
until their surfaces come into contact. The described behavior is the answer
to question B.

So far, we have approached the solution of the problem, but some questions
remain unanswered. For instance, we still have to justify our neglect of the
red curve and also explain the absence of equilibrium distances for $q>q_{cr}%
$. This will be done in the next section with the assistance of an energy function.

\section{\textbf{Energy and Stability}}

To understand the difference between the locally stable (green, upper branch
in \textit{Fig.2}) and unstable (red, lower branch) equilibrium states better,
it is useful to analyze the energy profiles of the system. Given that the
string is not stretchable, the potential energy $U$ of the system consists
only of Coulombic and gravitational components. One can express the total
potential energy as $U=2mgl~[1-\cos(\alpha)]-kq^{2}/s$. \ Recalling that
$\alpha$ is small and using the approximations $\cos(\alpha)\symbol{126}%
1-\alpha^{2}/2~$and$\ \alpha\symbol{126}~\sin(\alpha)=(s-s_{0})/2l$, we get

\begin{center}

\begin{equation}
U(s,q)=\frac{mg}{4l}(s-s_{0})-k\frac{q^{2}}{s} \label{U(s,q)}%
\end{equation}
\end{center}

\begin{center}

\begin{figure}

\begin{center}
\includegraphics[
height=2.9827in,
width=3.4904in
]%
{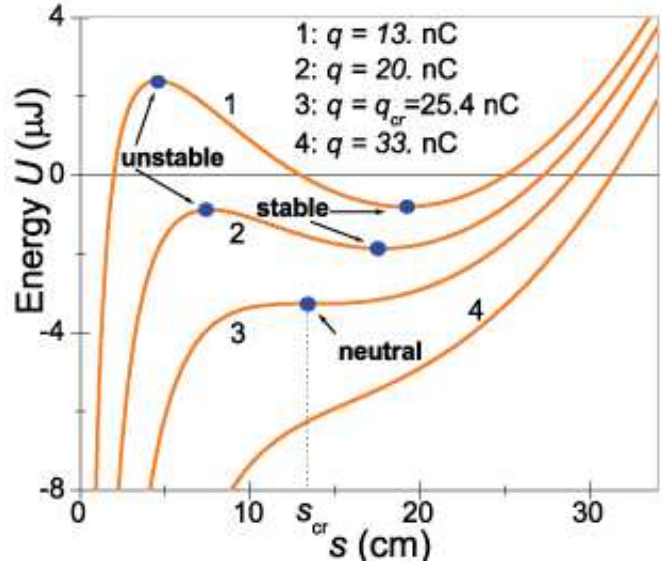}%
\end{center}
\caption{Energy profiles for different values of $q$.}


\end{figure}

\end{center}
Using the numeric values of the constants, and choosing representative values
of $q$, we plot the energy profiles $U(s,q)$ in\textit{ Fig.3.} For
$0~\mathit{<q<q}_{cr}$\textit{, }$\mathit{U(s,q)}$\textit{ }has a maximum at
smaller $s$ and a minimum at larger $s$. The points of minima describe locally
stable equilibrium states because small disturbances result in a restoring
force. In contrast, maxima points are unstable because the system moves away
from equilibrium after a small disturbance. This explains why the red branch
was discarded in the preceding discussion.

For $q=q_{cr}$ both solutions merge forming the inflection point, also termed
neutral equilibrium. The equilibrium disappears entirely for $q>q_{cr}$
because corresponding energy curves do not have a minima. (See curve 4 in
\textit{Fig.3.}) What is the cause of this anomaly? Let us note the different character of the
forces. The horizontal component of tension varies almost linearly with
$\alpha$ and correspondingly with $s$, whereas the Coulomb force is
proportional to the inverse square of the separation $s$. The Coulomb force
tends steeply to infinity as separation distance goes to 0. Consequently, for
every $\mathit{q\neq0}$ there exists a separation below which the Coulombic
attraction will always overwhelm the counteracting tension force. This
\textquotedblleft equiforce\textquotedblright\ boundary separation coincides
with the points of energy maxima and provides a physical interpretation of the
red branch.

Also note, as $q$ grows, the region where the electric force overwhelms the
horizontal tension widens until it engulfs the entire range at some value of
$q$ which is exactly the point $q_{cr}$. Above that, the only possible
equilibrium corresponds to the collapsed state. So it is this difference
between the behaviors of the two forces that causes the catastrophe.

\section{\textbf{Irreversibility and Hysteresis}}

We still did not answer the question C of the problem. So far we have been
assuming that $q$ was incremented starting from an initial value of 0 and
separation $s_{0}$. But what if we reverse the process by starting from the
collapsed state at charge value \textit{q $\geq$ q}$_{cr}$, and then
decremented $\mathit{q}$\textit{ to }$\mathit{q=0}$? At what point would the
charges separate?

This question has already been addressed for the case of point charges, where 
complete discharge is required for separation. We have also mentioned that
when the shell radius $R$ $\geq s_{cr}/2$, the catastrophe and corresponding
collapse disappear.

Thus, it only remains to consider radii $0$ $<$ $R$ $<s_{cr}/2$. \ Let's pick
a representative value of $~R\prime=2~$cm and repeat the analysis of energy
profiles familiar from the previous section. Note, the equilibrium separation
in the collapsed state is no longer determined by the original condition
$\overrightarrow{F}_{total}\ =\vec{T}+m\vec{g}+\vec{F}_{c}=0\ $leading to
\textit{Eq.3}. Rather, the non-zero force $\overrightarrow{F}_{total}$
attracting the charges is balanced here by the repelling force due to rigidity
of the shells.

\begin{center}%
\begin{figure}
[ptb]
\begin{center}
\includegraphics[
natheight=1.329200in,
natwidth=1.653500in,
height=2.7562in,
width=3.4878in
]
{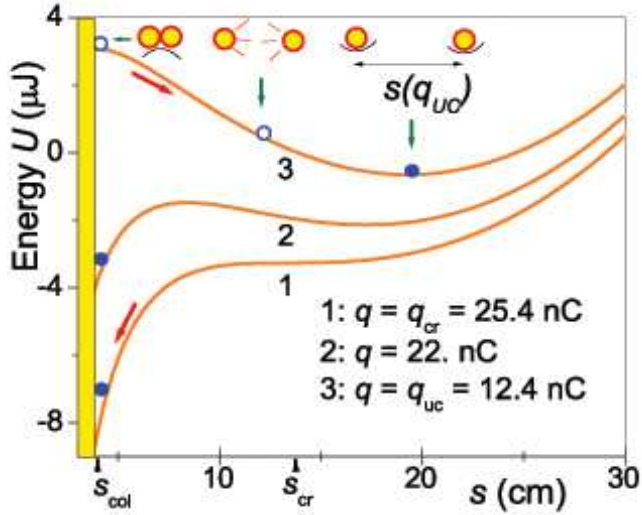}%
\end{center}
\caption{Presented are snapshots of energy profiles in
three progressive stages of discharge. The plots are restricted to the
accessible region \textit{s $\geq$ s}$_{col}$ . Initially, the system is in
the collapsed state and at charge $q_{cr}$. The yellow bar represents the hard
shell repulsion maintaining a lower bound on the separation distance at
$s_{col}=2R\prime=4$ cm. Blue dot is the proverbial ball residing in a
potential well. The cartoon depicts the uncollapse which happens when q
approaches $q_{uc}$.}

\end{figure}
\end{center}

Consequently, the condition for separation is when the net attractive force
$\overrightarrow{F}_{total}$ pressing the shells against each other vanishes.
Obviously, this condition leads exactly to \textit{Eq.3}. However, now it
should be treated differently. Instead of finding an equilibrium separation
for a fixed $q$, we use it to find a $q\ $value for which the force at a fixed
separation $s=2R~$vanishes.

For our example case of $s_{col}$ $=4\ $ cm, the $q_{uc}$ determined by
\textit{Eq.3} corresponds to the intersection point between the horizontal
line $s=s_{col}$ and the curve $s(q)$ (see Fig. 2, point d ).

This result can be explained using the energy profiles in \textit{Fig. 4}.
Profile 1 corresponds to the collapsed state at $q=q_{cr}$. The attraction
between the charges is compensated by contact repulsive force (see the yellow
wall). The dip in the energy profile at distances close to $s_{col}$, creates
an effective energy well which stabilizes the system.

Profile 2 illustrates how a decrease in the charge to $22$ nC makes the energy
well trapping the system more shallow while still keeping the charges together.

Reducing $q$ further eventually leads to the scenario of profile 3. When
$q_{uc} \sim 12.4$ nC is reached, the slope of the energy function at
$s_{col}$ becomes horizontal. This means that the attractive force keeping the
spheres together vanishes and the charge finds itself on the top of an energy
hill. In other words, the equilibrium becomes unstable, so the charge rolls
down to the stable energy well. This transition corresponds to the vertical
line $d^{\prime}b~$connecting the red (maximum) and green (minimum)
equilibrium positions in \textit{Fig.2}. Since it is sudden, we term it an
\textit{uncollapse \footnote{In fact the uncollapse can occur before $q_{uc}$
is reached. As we learned from energy profiles (\textit{Figs.3\& 4}) for every
$q$ between $q_{col}$ and $q_{cr}$ there exists a second energy well
corresponding to the locally stable equilibrium distance expressed by the
green branch of \textit{Fig.2 }and separated from the collapsed state by the
presence of a barrier. This is an analog of the chemical activation energy
barrier useful for understanding various molecular processes. A sufficient
perturbation due to thermal, mechanical or electrical fluctuations could knock
the system out of the collapsed state and into this stable well. In general,
the fluctuations tend to narrow the hysterisis.}.}
\textit{Fig.2} outlines the charging cycle for R = 2 cm with the path
\textit{abcd'da}. The vertical segment \textit{cd'} corresponds to the
collapse, when separation suddenly changes from $13.4~$to $4$ cm. Additional
charging will not change the separation. Decrementing $q$ after the collapse
initially keeps the charges at fixed separation 4 cm, until $q_{ucl}\sim
12.4$ nC is reached (segment $d^{\prime}d)$. After that point, $s$ undergoes a
vertical transition, an uncollapse, from the red to the green branch,
\textit{db}. Further decreases in $q$ will move charges apart along the green branch.

To extract a general principle from this example, we note that in the presence
of catastrophe, the forward (charging) and reverse (discharging) behaviors are
different. This irreversible behavior is called hysteresis. It is very common
in many physical phenomena associated with instabilities, phase transitions
and catastrophes.

\section{Conclusion}

The solution of the conventional problem of identical suspended charges $q$
describes a continuous relation between the equilibrium state and the value of
$q$. Increasing $q$ leads to a monotonic increase of the deflection
angle$\ \alpha$. For large values of $q$ the angle asymptotically approaches
$\pi/2$ and the separation $s$ correspondingly approaches $2l$. Such behavior,
when equilibrium properties smoothly depend on external parameters (pressure,
charge, temperature, etc.) can be considered \textit{normal}%
$\ \cite{ParPar2004}~$and is typical for practically all the equilibrium
problems encountered in high school physics.

However, the real world is full of examples when smooth variation of external
parameters results in sudden catastrophic change of equilibrium properties.
The colloquialism, ``the straw that broke the camel's back'', expresses
exactly this idea. These sorts of phenomena are directly related to physical
catastrophes, instabilities and phase transitions.

As we can see, a generalization of the aforementioned problem for the case of
opposite charges immediately results in catastrophic behavior. At a certain
value of $q$ the equilibrium suddenly disappears, and system undergoes a sharp
and discontinuous transition to a new equilibrium state. Usually, analysis of
the catastrophic behavior requires very complex physics and mathematics. The
discussion presented here and in references \cite{ParPar2004,Par2002}
demonstrates that in some cases it can be accomplished with analytical tools
available to high school physics students.

In this paper, we have also introduced the concept of hysteresis which is
another feature of catastrophic behavior. It was shown that equilibrium
separation between the charges for a given $q$ value can depend on charging
history. We demonstrated that in some cases the same value of $q$, depending
on how it was reached, can correspond either to a comparatively large
separation or to a collapsed state where charges are stuck together. We
believe guided investigation of the aforementioned phenomena can result in
interesting student research projects. \bigskip

\bibliographystyle{plain}
\bibliography{ref}

\end{document}